# System Reliability, Fault Tolerance and Design Metrics Tradeoffs in the Distributed Minority and Majority Voting Based Redundancy Scheme

'""""""""""""""""""""P. BALASUBRAMANIAN*
* School of Computer Engineering
Nanyang Technological University
50 Nanyang Avenue
SINGAPORE 639798
E-mail: balasubramanian@ntu.edu.sg

*Abstract:* - The distributed minority and majority voting based redundancy (DMMR) scheme was recently proposed as an efficient alternative to the conventional N-modular redundancy (NMR) scheme for the physical design of mission/safety-critical circuits and systems. The DMMR scheme enables significant improvements in fault tolerance and design metrics compared to the NMR scheme albeit at the expense of a slight decrease in the system reliability. In this context, this paper studies the system reliability, fault tolerance and design metrics tradeoffs in the DMMR scheme compared to the NMR scheme when the majority logic group of the DMMR scheme is increased in size relative to the minority logic group. Some example DMMR and NMR systems were realized using a 32/28nm CMOS process and compared. The results show that 5-of-M DMMR systems have a similar or better fault tolerance whilst requiring similar or fewer function modules than their counterpart NMR systems and simultaneously achieve optimizations in design metrics. Nevertheless, 3-of-M DMMR systems have the upper hand with respect to fault tolerance and design metrics optimizations than the comparable NMR and 5-of-M DMMR systems. With regard to system reliability, NMR systems are closely followed by 5-of-M DMMR systems which are closely followed by 3-of-M DMMR systems. The verdict is 3-of-M DMMR systems are preferable to implement higher levels of redundancy from a combined system reliability, fault tolerance and design metrics perspective to realize mission/safety-critical circuits and systems.

## 1 Introduction to NMR and DMMR

N-modular redundancy (NMR) has been widely used for the fault-tolerant design of mission and safety-critical circuits and systems [1] – [3] which are used in space, aerospace, nuclear, defence, banking, financial and other industrial applications. Reference [4] suggests that due to increasing reliability and variability issues in the nanoelectronics regime, a selective utilization of higher levels of redundancy which involves the use of several identical function modules called as progressive modular redundancy may be needed for realizing specific portions of future generation mission and safety-critical circuits and systems which demand greater fault tolerance. In this context, it was recently shown [5] that the distributed minority and majority voting based redundancy (DMMR) scheme forms an efficient and viable alternative to the conventional NMR scheme for implementing mission and safety-critical circuits and systems.

In the NMR scheme, N identical function modules are used and the equivalent outputs of N identical function modules are supplied to a majority voter, which processes and produces the NMR system output by performing majority voting on the function modules outputs. The block schematic of the NMR system is shown in Fig 1. In the NMR scheme, N is odd, and at least $(N+1)/2$ function modules are required to satisfy the majority logic, i.e. the faulty/failure state of at most $(N-1)/2$ function modules are tolerated by the NMR scheme.

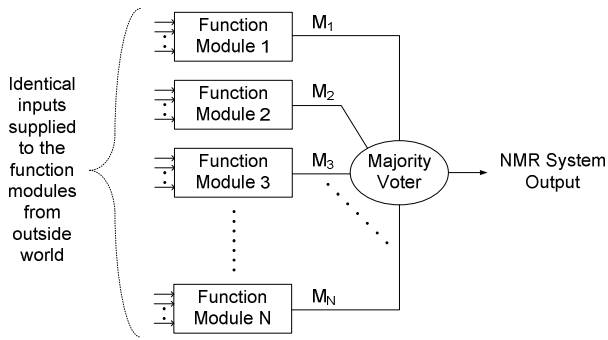

Fig 1. Block diagram of NMR system

In the DMMR scheme, the function modules are split into two groups as the majority logic group and the minority logic group. A majority of function modules in the majority logic group and at least one function module in the minority logic group should maintain the correct operation to ensure the correct operation of the DMMR system. Note that in a generic K-of-M DMMR system, K identical function modules constitute the majority logic group and the remaining (M−K) identical function modules constitute the minority logic group, where K and M are positive integers with $K < M$ and $K \geq 3$. The equivalent outputs produced by identical function modules of the majority and minority logic groups are combined by the DMMR voter, as shown in Fig 2, to produce the DMMR system output.

## 2  3-of-M and 5-of-M DMMR Systems

Fig 2a shows the 3-of-M DMMR system topology where the majority logic group comprises 3 function modules and the minority logic group comprises (M–3) function modules. Fig 2b shows the 5-of-M DMMR system topology where the majority logic group comprises 5 function modules and the minority logic group comprises (M–5) function modules. The corresponding DMMR voters are also shown in Fig 2. The majority voting element present in the DMMR voter of the 3-of-M DMMR system is a 2-of-3 majority voter [6], and the majority voting element present in the DMMR voter of the 5-of-M DMMR system is a 3-of-5 majority voter [7].

Referring to Fig 2a, in the 3-of-M DMMR system, at least two function modules out of $F_1$, $F_2$, $F_3$ comprising the majority logic group should maintain the correct operation and at least one function module amongst $F_4$ to $F_M$ in the minority logic group should maintain the correct operation concurrently. On the other hand, referring to Fig 2b, in the 5-of-M DMMR system, at least three function modules out of $F_1$, $F_2$, $F_3$, $F_4$, $F_5$ comprising the majority logic group should maintain the correct operation and at least one function module amongst $F_6$ to $F_M$ in the minority logic group should maintain the correct operation concurrently. In this article, we succinctly study the system reliability, fault tolerance and design metrics aspects of some example 5-of-M DMMR systems in comparison with those of counterpart NMR and 3-of-M DMMR systems to understand the tradeoffs involved.

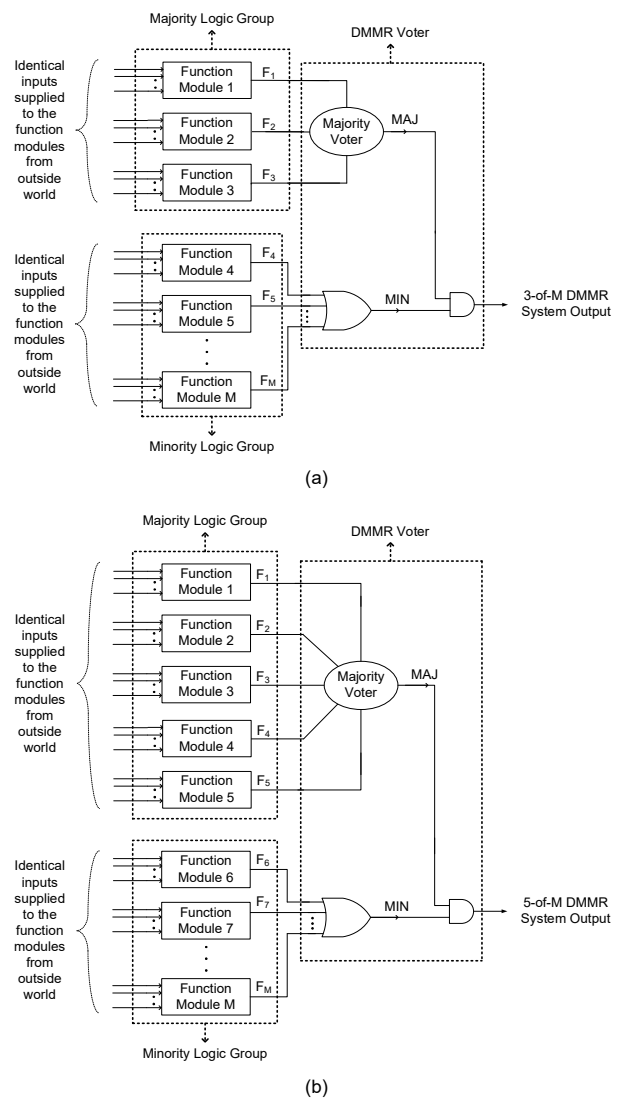

Fig 2. Block diagrams of (a) 3-of-M DMMR system and (b) 5-of-M DMMR system

Equations (1) and (2) give the respective system reliability expressions of 5-of-7 and 5-of-8 DMMR systems by assuming perfect DMMR voters. The system reliability equations of counterpart NMR (i.e. 7MR and 9MR) and 3-of-M DMMR (i.e. 3-of-6 DMMR and 3-of-7 DMMR) systems are given in [5]. In (1) and (2), $R_M$ denotes the module reliability and $R_S$ denotes the system reliability. Since identical function modules are used, the modules reliabilities are assumed to be equivalent. $R_M$ denotes the reliability or the probability of correct working of a function module, and $(1-R_M)$ denotes the non-reliability or probability of the faulty/failure state of a function module.

$$R_S^{\text{5-of-7 DMMR}} = 20R_M^4(1-R_M)^3 + 20R_M^5(1-R_M)^2 + 7R_M^6(1-R_M) + R_M^7 \quad (1)$$

$$R_S^{\text{5-of-8 DMMR}} = 30R_M^4(1-R_M)^4 + 45R_M^5(1-R_M)^3 + 28R_M^6(1-R_M)^2 + 8R_M^7(1-R_M) + R_M^8 \quad (2)$$

7MR, 3-of-6 DMMR and 5-of-7 DMMR systems would be able to accommodate the faulty or failure state of a maximum of 3 function modules, while 9MR, 3-of-7 DMMR and 5-of-8 DMMR can cope with the faulty or failure state of a maximum of 4 function modules. With respect to the former, the 3-of-6 DMMR system requires one function module less than the 7MR and 5-of-7 DMMR systems. With respect to the latter, the 5-of-8 DMMR system requires one function module less than the 9MR system and one function module more than the 3-of-7 DMMR system.

An analysis of the system reliabilities of 7MR, 9MR, 3-of-6 DMMR, 3-of-7 DMMR, 5-of-7 DMMR and 5-of-8 DMMR systems versus their module reliabilities reveals that NMR systems feature the highest reliability closely followed by the reliabilities of 5-of-M DMMR systems which are closely followed by the reliabilities of 3-of-M DMMR systems. For example, for $R_M = 0.9$, the system reliabilities of 7MR and 9MR systems are 0.997272 and 0.99910908; the system reliabilities of 5-of-7 and 5-of-8 DMMR systems are 0.9815256 and 0.99044856, and the system reliabilities of 3-of-6 and 3-of-7 DMMR systems are 0.971028 and 0.9719028 respectively. In general, 5-of-M DMMR systems show improvements in system reliability compared to the system reliabilities of 3-of-M DMMR systems whilst featuring the same degree of fault tolerance although requiring a similar or greater number of function modules. Hence, the system reliabilities of 5-of-M DMMR systems happen to lie midway between the corresponding system reliabilities of counterpart NMR and 3-of-M DMMR systems.

## 3 Simulation Results and Conclusions

Some example NMR, 3-of-M DMMR and 5-of-M DMMR systems especially targeting higher levels of redundancy were implemented based on a 32/28nm CMOS process [8] by considering the 4×4 array multiplier [9] as a representative function module as in [5]. The structural integrity of the different redundant systems was preserved during technology mapping to facilitate a legitimate comparison of their corresponding design metrics post-physical synthesis. The simulation mechanism, environment (typical case PVT) and test benches (supplied at time intervals of 4ns, i.e. 250MHz) were maintained the same as that of [5] to ensure uniformity and to enable a direct correspondence with the previous results. The simulation results viz. average power, critical path delay, and Silicon area of the different redundant systems estimated are given in Table 1. A figure-of-merit (FOM) was calculated as the inverse of the power-delay-area product as in [5], which is also given in Table 1. Since power, delay and area are desirable to be minimized, a high value of FOM indicates an optimized design [10] – [17], which signifies reduction in cost.

Table 1. Power, delay and area metrics of various NMR, 3-of-M DMMR and 5-of-M DMMR systems

| NMR/DMMR System Type | Power (µW) | Delay (ns) | Area (µm²) | FOM (× 10⁶) |
|---|---|---|---|---|
| 7MR | 191.2 | 1.12 | 865.11 | 5.40 |
| 3-of-6 DMMR | 129.4 | 0.90 | 567.25 | 15.14 |
| 5-of-7 DMMR | 164.1 | 0.99 | 730.92 | 8.42 |
| 9MR | 278.5 | 1.23 | 1269.7 | 2.30 |
| 3-of-7 DMMR | 151.2 | 0.91 | 661.79 | 10.98 |
| 5-of-8 DMMR | 184.5 | 0.99 | 817.33 | 6.70 |

Compared to 7MR, the 3-of-6 DMMR and 5-of-7 DMMR systems implementations report respective improvements in FOM by 180.4% and 56%. In comparison with 9MR, the 3-of-7 DMMR and 5-of-8 DMMR systems implementations report corresponding enhancements in FOM by 377.4% and 191.3%. The simulation results show that the FOM of 5-of-M DMMR systems lies approximately midway between the FOMs of counterpart NMR and 3-of-M DMMR systems.

Hence it can be inferred that when giving importance to system reliability alone, the NMR system topology is preferable. However, to realize higher levels of redundancy, the NMR scheme would be a poor choice when taking into account the system implementation cost. In comparison, the 5-of-M DMMR system topology would be a better choice as it could considerably reduce the implementation cost whilst featuring only slightly less system reliability. Nevertheless, from a combined system reliability, fault tolerance and design metrics (i.e. cost and weight) perspective, the 3-of-M DMMR system is all the more preferable than the rest whilst being associated with a moderately less system reliability.